\documentclass[a4paper]{jpconf}
\usepackage{graphicx}
\begin{document}
\title{Correlations between charge and heat currents\\in an interacting quantum dot}

\author{Adeline Cr\'epieux$^1$, Paul Eym\'eoud$^1$ and Fabienne Michelini$^2$}

\address{$^1$ Aix Marseille Universit\'e, Universit\'e de Toulon, CNRS, CPT, UMR 7332, 13288 Marseille, France}
\address{$^2$ Aix Marseille Universit\'e, CNRS, IM2NP, UMR 7334, 13288 Marseille, France}

\ead{adeline.crepieux@cpt.univ-mrs.fr}

\begin{abstract}
We consider an interacting quantum dot connected to two reservoirs driven at distinct voltage/temperature and we study the correlations between charge and heat currents first as a function of the applied voltage bias, and second as a function of the temperature gradient between the two reservoirs. The Coulomb interactions in the quantum dot are treated using the Hartree approximation and the dot occupation number is determined self-consistently. The correlators exhibit structures in their voltage dependency which are highly non-linear when the coupling between the dot and the reservoirs is weak, and their behavior with temperature is non-monotonous. Moreover the sign of heat cross-correlator can change contrary to what happens with the charge cross-correlator which is always negative. 
The presence of Coulomb interactions enlarges the domain of voltage in which the heat cross-correlator is negative.
%The presence of Coulomb interactions in the dot does not change these features.
\end{abstract}

\section{Introduction}

The field of quantum thermoelectricity is very active right now. Its main objectives are the increase of the thermoelectric conversion efficiency by reducing the size of the system and the possibility to build on-chip thermoelectric nanodevices. Among the recent experimental works, we can cite the measurement of non-linear thermovoltage and thermocurrent in quantum dots~\cite{fahlvik2013}, the study  of the Seebeck effect in magnetic tunnel junctions~\cite{walter2011} and the measurement of first, second and third harmonic voltage responses in nanoscale spin valves~\cite{bakker2010}. In parallel, theoretical activities are needed. The purpose here is to understand how Coulomb interactions in the quantum dot affect the profile of the different kinds of currents correlators that one can define when both charge and heat fluxes are present, i.e., the charge/charge correlator, the heat/heat correlator as well as the mixed charge/heat correlator. This last quantity has been introduced quite recently~\cite{giazotto2006} and has been considered so far only in a three-terminal thermoelectric device with two quantum dots in the Coulomb blockade regime~\cite{sanchez2013}. It deserves to be studied in a more systematic way since it has been shown that it allows to quantify the thermoelectric conversion in both linear response regime and Schottky regime~\cite{crepieux2014}.

\section{Model and method}

The starting point to describe the interacting quantum dot is the Anderson model~\cite{anderson1961}:
\begin{eqnarray}\label{H}
H=\sum_{\sigma,k\in\{L,R\}}\epsilon_k \hat c^\dag_{k\sigma}\hat c_{k\sigma}
+\sum_\sigma\epsilon_0  d^\dag_{\sigma}\hat d_{\sigma}+U  d^\dag_{\uparrow}\hat d_{\uparrow}\hat d^\dag_{\downarrow}\hat d_{\downarrow}+\sum_{\sigma,k\in\{L,R\}} t_k\hat c^\dag_{k\sigma}\hat d_{\sigma}+h.c.~,
\end{eqnarray}
where $\hat d^\dag_{\sigma}$ ($\hat d_{\sigma}$) and $\hat c^\dag_{k\sigma}$ ($\hat c_{k\sigma}$) are the creation (annihilation) operators of one electron with spin $\sigma$ in the dot and in the reservoirs respectively. $\epsilon_0$ is the energy level of the dot and $U$ measures the Coulomb interactions strength. $\epsilon_k$ is the energy band dispersion assumed to be identical in the left (L) and right (R) reservoirs and $t_k$ is the hopping amplitude between the reservoirs and the dot.

\begin{figure}
\begin{center}
\includegraphics[width=20pc]{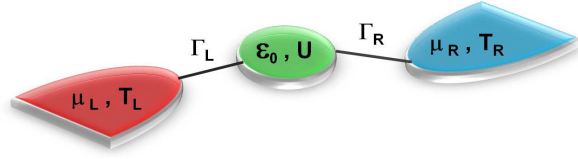}
\caption{\label{schema}Schematic picture of the single level interacting quantum dot connected to left~(L) and right~(R) reservoirs driven at distinct chemical potentials $\mu_{L,R}$ and temperatures $T_{L,R}$. The left and right barriers are assumed to be identical: $\Gamma_L=\Gamma_R$, and the convention chosen for the currents direction is from the reservoirs to the dot.}
\end{center}
\end{figure}

We calculate the zero-frequency Fourier transforms of the correlations between the charge current $\hat I_p^{e}=-\sum_{\sigma,k\in p} \dot N_{k\sigma}/e$, with $N_{k\sigma}=\hat c^\dag_{k\sigma}\hat c_{k\sigma}$, and the heat current $\hat I_p^{h}=-\sum_{\sigma,k\in p} (\epsilon_k-\mu_p)\dot N_{k\sigma} $, in $p$ and $q$ reservoirs:
\begin{eqnarray}
\mathcal{S}^{ee}_{pq}&=&\int^{\infty}_{-\infty} \langle\delta \hat I^e_{p}(t)\delta\hat I^e_{q}(0)\rangle dt~,\;\;\;\;\;\;\;\;\;\mathcal{S}^{hh}_{pq}=\int^{\infty}_{-\infty} \langle\delta \hat I^h_{p}(t)\delta\hat I^h_{q}(0)\rangle dt~,\\
\mathcal{S}^{eh}_{pq}&=&\int^{\infty}_{-\infty} \langle\delta \hat I^e_{p}(t)\delta\hat I^h_{q}(0)\rangle dt~,\;\;\;\;\;
\;\;\;\;\mathcal{S}^{he}_{pq}=\int^{\infty}_{-\infty} \langle\delta \hat I^h_{p}(t)\delta\hat I^e_{q}(0)\rangle dt~,
\end{eqnarray}
where $\delta\hat I^{e(h)}_{p}(t)=\hat I^{e(h)}_{p}(t)-\langle \hat{I}^{e(h)}_p\rangle$. $\mathcal{S}^{ee}_{pq}$ corresponds to the charge correlator, $\mathcal{S}^{hh}_{pq}$ corresponds to the heat correlator, and we call $\mathcal{S}^{eh}_{pq}$ ($\mathcal{S}^{he}_{pq}$) the mixed correlator which measures the correlation between the charge (heat) current in reservoir $p$ and the heat (charge) current in reservoir $q$.

Using the Landauer-B\"uttiker formalism~\cite{blanter2000}, the auto-correlators read as \cite{crepieux2014}:
\begin{eqnarray}\label{Seepp}
\mathcal{S}^{ee}_{pp}&=&\frac{e^2}{h}\int^{\infty}_{-\infty}\mathcal{F}(\epsilon)d\epsilon~,
\end{eqnarray}
\begin{eqnarray}\label{Sehpp}
\mathcal{S}^{eh}_{pp}&=&\mathcal{S}^{he}_{pp}=\frac{e}{h}\int^{\infty}_{-\infty}(\epsilon-\mu_{p})\mathcal{F}(\epsilon)d\epsilon~,
\end{eqnarray}
\begin{eqnarray}
\mathcal{S}^{hh}_{pp}&=&\frac{1}{h}\int^{\infty}_{-\infty}(\epsilon-\mu_{p})^2\mathcal{F}(\epsilon)d\epsilon~,
\end{eqnarray}
where
\begin{eqnarray}\label{F}
\mathcal{F}(\epsilon)&=&\mathcal{T}(\epsilon)\bigg[f_L(\epsilon)[1-f_L(\epsilon)]+f_R(\epsilon)[1-f_R(\epsilon)]\bigg]
+\mathcal{T}(\epsilon)\Big[1-\mathcal{T}(\epsilon)\Big][f_L(\epsilon)-f_R(\epsilon)]^2~,
\end{eqnarray}
with $f_{L,R}(\epsilon)=[1+\exp((\epsilon-\mu_{L,R})/(k_BT_{L,R}))]^{-1}$ the Fermi-Dirac distribution function, and $\mathcal{T}(\epsilon)$ the transmission coefficient through the barriers which is assumed to be voltage independent. The cross-correlators read as:
\begin{eqnarray}\label{Seepbarp}
\mathcal{S}^{ee}_{p\bar{p}}&=&-\frac{e^2}{h}\int^{\infty}_{-\infty}\mathcal{F}(\epsilon)d\epsilon~,
\end{eqnarray}
\begin{eqnarray}
\mathcal{S}^{eh}_{p\bar{p}}&=&-\frac{e}{h}\int^{\infty}_{-\infty}(\epsilon-\mu_{\bar{p}})\mathcal{F}(\epsilon)d\epsilon~,
\end{eqnarray}
\begin{eqnarray}
\mathcal{S}^{he}_{p\bar{p}}&=&-\frac{e}{h}\int^{\infty}_{-\infty}(\epsilon-\mu_{p})\mathcal{F}(\epsilon)d\epsilon~,
\end{eqnarray}
\begin{eqnarray}\label{Shhpbarp}
\mathcal{S}^{hh}_{p\bar{p}}&=&-\frac{1}{h}\int^{\infty}_{-\infty}(\epsilon-\mu_{p})(\epsilon-\mu_{\bar{p}})\mathcal{F}(\epsilon)d\epsilon~,
\end{eqnarray}
where $\bar{p}=R$ when $p=L$, and $\bar{p}=L$ when $p=R$.

To treat the Coulomb term of Eq.~(\ref{H}), we use the Hartree approximation for which the retarded Green's function associated to the dot is given by~\cite{mahan90}: $G^r_\sigma(\epsilon)=[\epsilon-\epsilon_0-U\langle n_{\bar\sigma}\rangle-\Sigma_\sigma(\epsilon)]^{-1}$. The self-energy is defined as $\Sigma_\sigma(\epsilon)=-i\Gamma_\sigma(\epsilon)$, where $\Gamma_\sigma(\epsilon)=2\pi\sum_{p\in \{L,R\}}|t_p|^2\rho_{p\sigma}(\epsilon)$ with $\rho_{p\sigma}$ the density of states for $\sigma$ spin in the reservoir $p$. We use the notation ${\bar\sigma}=\downarrow$ when $\sigma=\uparrow$, and ${\bar\sigma}=\uparrow$ when $\sigma=\downarrow$. The average dot occupation numbers for spin up and spin down are $\langle n_{\uparrow}\rangle=\langle \hat d^\dag_{\uparrow} \hat d_{\uparrow}\rangle$ and $\langle n_{\downarrow}\rangle=\langle \hat d^\dag_{\downarrow} \hat d_{\downarrow}\rangle$ respectively.  To determine these average dot occupation numbers, we use a self-consistent numerical resolution using the relation~\cite{ng1996}:
\begin{eqnarray}
\langle n_\sigma\rangle=-\sum_{p=L,R}\int^{\infty}_{-\infty} \frac{d\epsilon}{2\pi}f_p(\epsilon)\mathrm{Im}\{G^r_\sigma(\epsilon)\}~,
\end{eqnarray}
assuming that the density of states is spin and energy independent (non magnetic reservoirs in the wide-band limit). Once the dot occupations are obtained, we calculate the auto-correlators and the cross-correlators using for the transmission coefficient the expression:
\begin{eqnarray}
\mathcal{T}(\epsilon)=\sum_\sigma\frac{\Gamma^2}{(\epsilon-\epsilon_0-U\langle n_\sigma\rangle)^2+\Gamma^2}~,
\end{eqnarray}
where we take $\Gamma=\Gamma_\uparrow=\Gamma_\downarrow$ (symmetrical barriers).

\section{Results}

In this section, we discuss the numerical results we have obtained for the profile of noises first as a function of the applied voltage bias: $V=(\mu_L-\mu_R)/e$, and second as a function of the temperature gradient: $T=T_L-T_R$. The other parameters are the average temperature of the system $T_0=(T_L+T_R)/2$, the dot energy level $\epsilon_0$, the Coulomb interactions strength $U$, and the coupling amplitude $\Gamma$ between the dot and reservoirs. We plot only the graphs for the auto-correlators $\mathcal{S}_{LL}^{ee}$, $\mathcal{S}_{LL}^{eh}$ and $\mathcal{S}_{LL}^{hh}$ since from Eq.~(\ref{Sehpp}), we see that $\mathcal{S}_{LL}^{eh}=\mathcal{S}_{LL}^{he}$. Similarly, we plot only the graphs for the cross-correlators $\mathcal{S}_{LR}^{eh}$, $\mathcal{S}_{LR}^{he}$ and $\mathcal{S}_{LR}^{hh}$ since from Eqs.~(\ref{Seepp}) and (\ref{Seepbarp}), we see that $\mathcal{S}_{LR}^{ee}=-\mathcal{S}_{LL}^{ee}$.

\begin{figure}
\includegraphics[width=12pc]{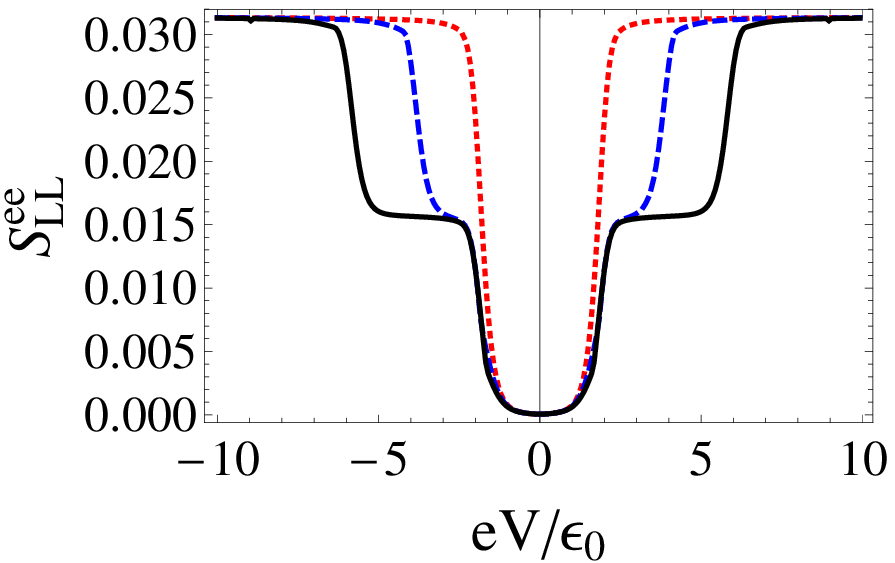}
\includegraphics[width=12pc]{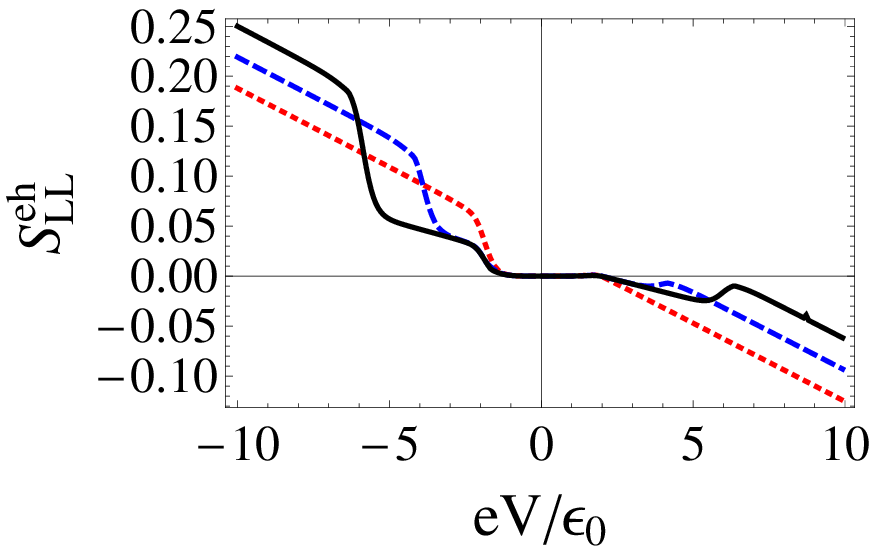}
\includegraphics[width=12pc]{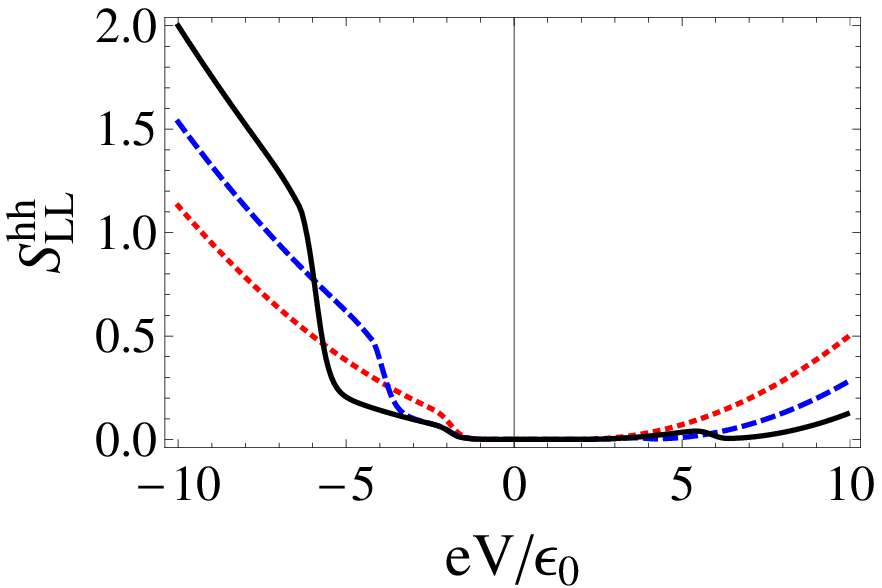}\\
\includegraphics[width=12pc]{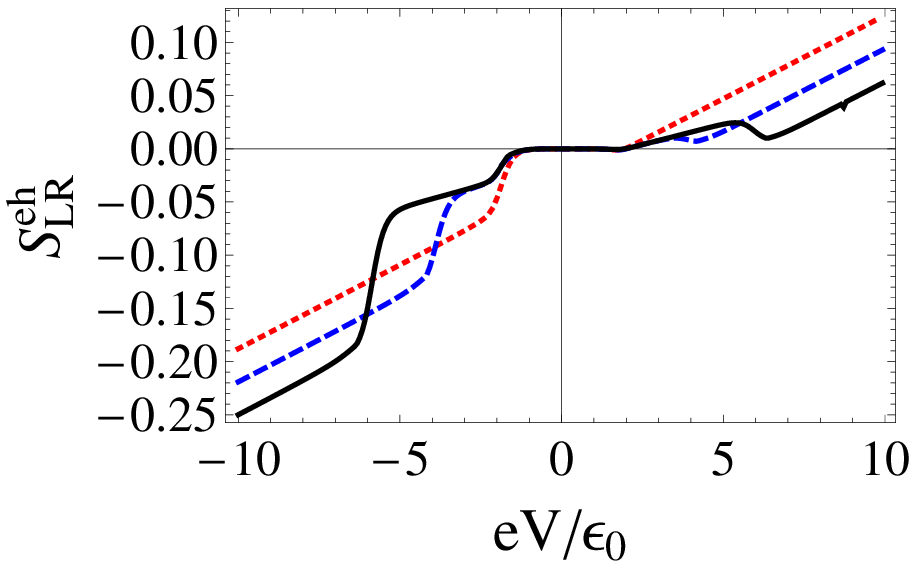}
\includegraphics[width=12pc]{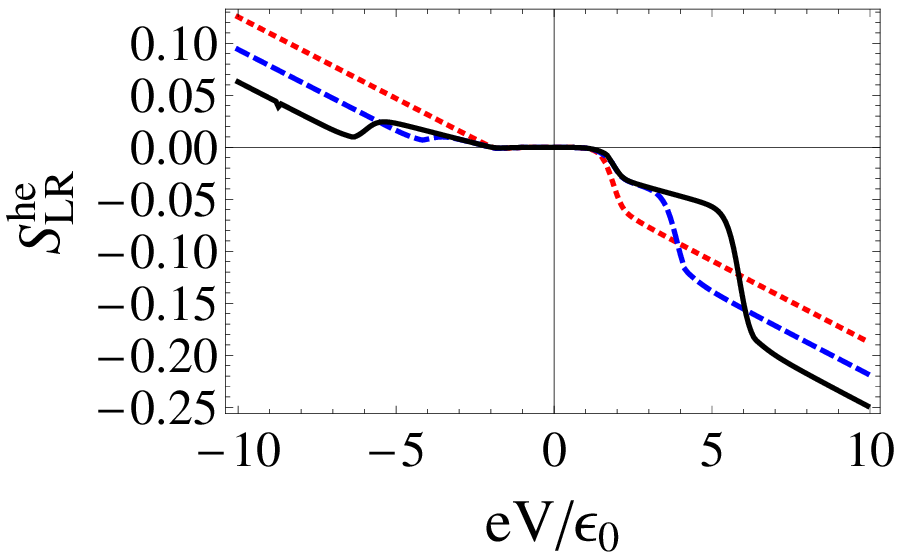}
\includegraphics[width=12pc]{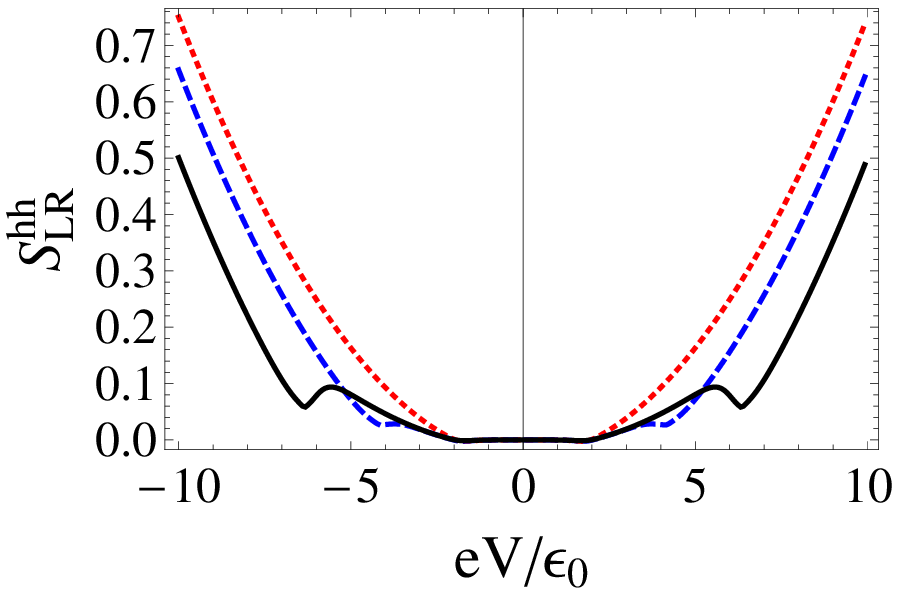}
\caption{\label{weak_gamma_V}Current correlators as a function of voltage at $\Gamma/\epsilon_0=0.01$, $T_0/\epsilon_0=0.1$, and $T=0$, for $U=0$ (red dotted lines), $U/\epsilon_0=1$ (blue dashed lines) and $U/\epsilon_0=2$ (black solid lines). The units for the correlators are $e^2\epsilon_0/h$ for $\mathcal{S}_{pq}^{ee}$, $\epsilon_0^3/h$ for $\mathcal{S}_{pq}^{hh}$, and $e\epsilon_0^2/h$ for $\mathcal{S}_{pq}^{eh}$ and $\mathcal{S}_{pq}^{he}$.}
\end{figure}

\begin{figure}
\includegraphics[width=12pc]{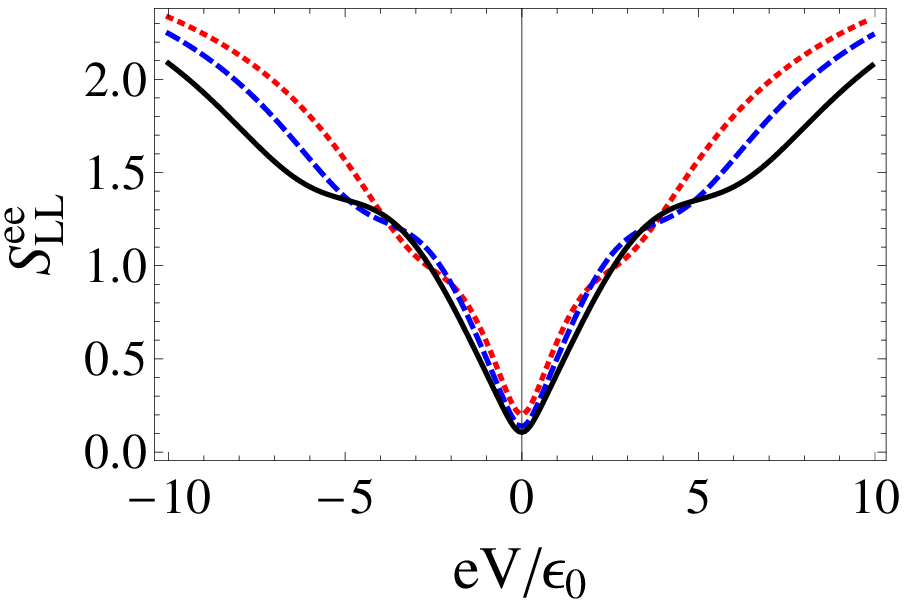}
\includegraphics[width=12pc]{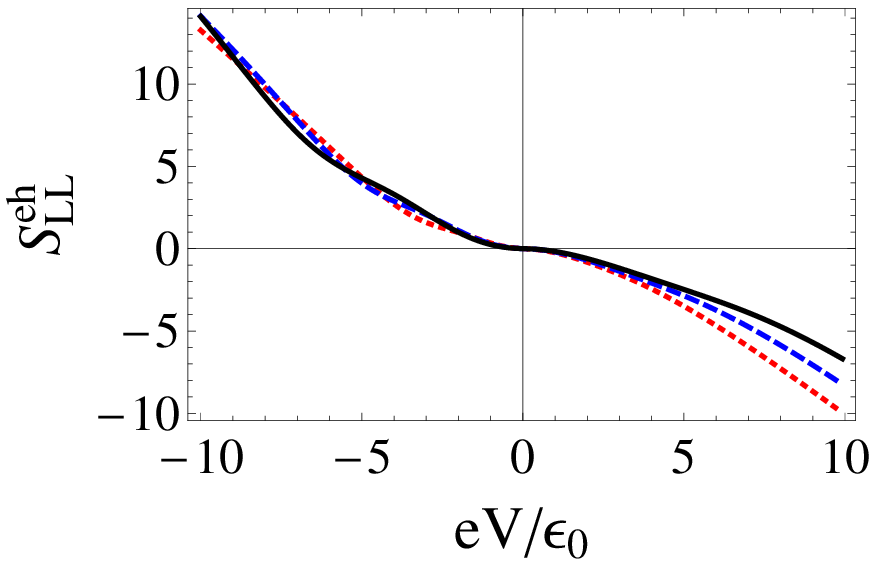}
\includegraphics[width=12pc]{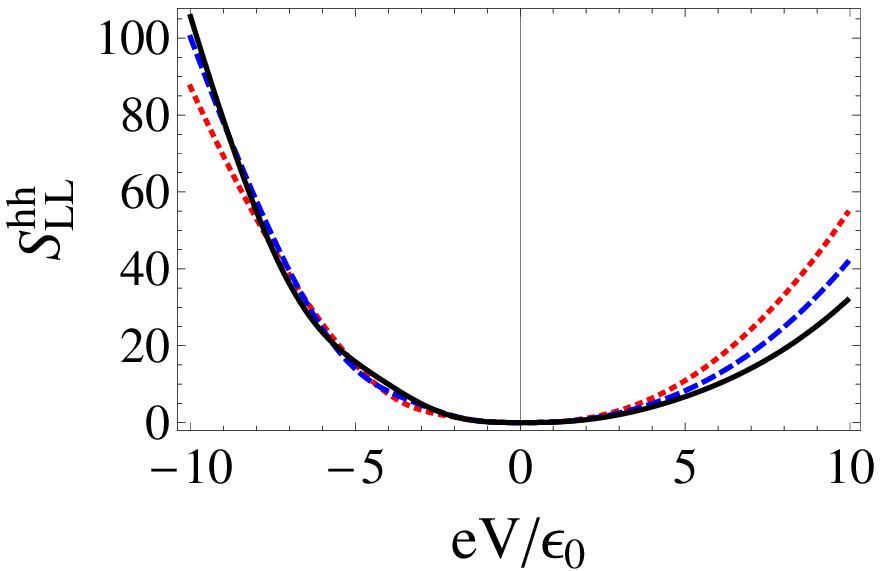}\\
\includegraphics[width=12pc]{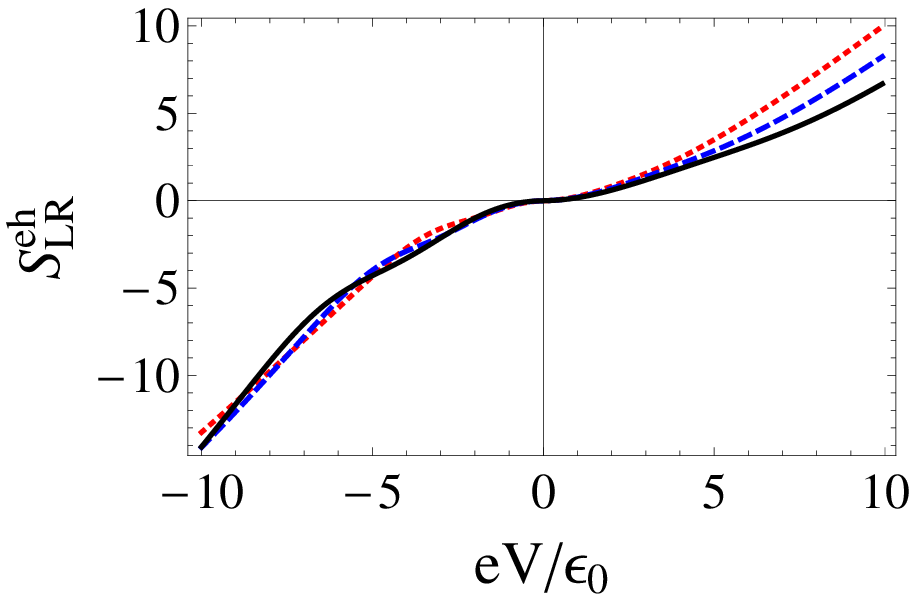}
\includegraphics[width=12pc]{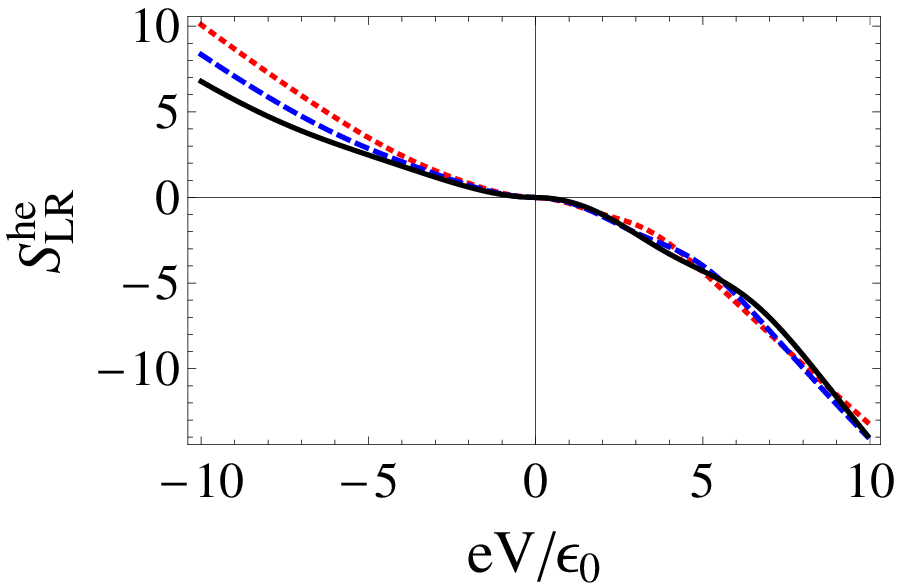}
\includegraphics[width=12pc]{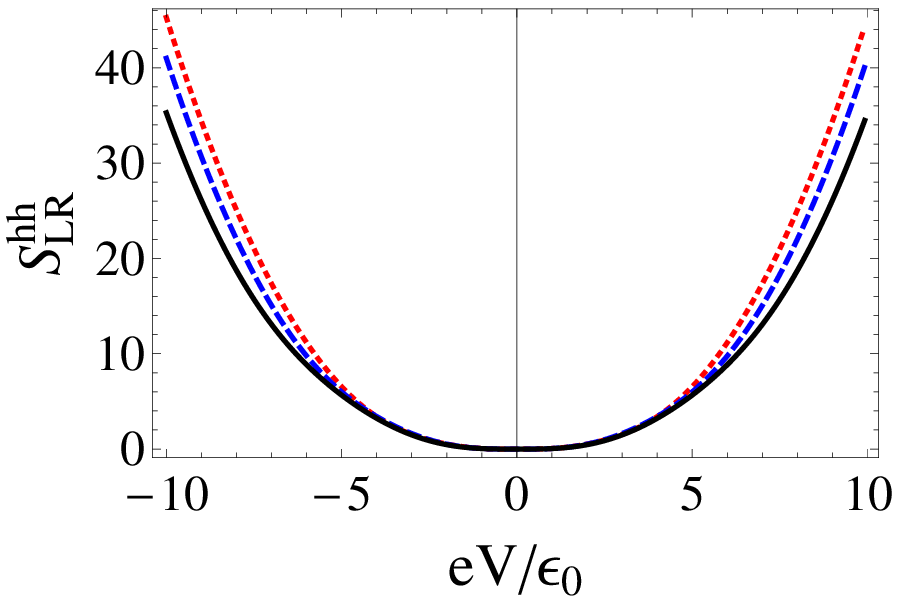}
\caption{\label{strong_gamma_V}Current correlators as a function of voltage at $\Gamma/\epsilon_0=1$, $T_0/\epsilon_0=0.1$, and $T=0$, for $U=0$ (red dotted lines), $U/\epsilon_0=1$ (blue dashed lines) and $U/\epsilon_0=2$ (black solid lines).}
\end{figure}

\subsection{Current correlators as a function of the bias voltage}

Figure~\ref{weak_gamma_V} shows the auto- and cross-correlators for a weak value of the coupling $\Gamma$  (Schottky limit) and increasing values of Coulomb interactions. In the absence of interaction (red dotted lines), the charge noise $\mathcal{S}_{LL}^{ee}$ has a single step, whereas in the presence of interactions (blue dashed and black solid lines), it exhibits two steps due to the transfer of a first electron at $eV/2\approx\pm \epsilon_0$, followed by the transfer of a second electron with an opposite spin at $eV/2\approx\pm(\epsilon_0+U)$. Indeed, in the Schottky regime, the charge noise is proportional to the absolute value of the charge current which has the same profile than the dot occupation number. All the other correlators exhibit also structures at  $eV/2\approx\pm\epsilon_0$ and  $eV/2\approx\pm(\epsilon_0+U)$ and have linear or quadratic variation in between these voltage values. 
At higher voltages and in the absence of interaction, the correlators vary as power laws with an exponent equals to zero for the charge correlators, equals to one for the mixed correlators and equals to two for the heat correlators (see the red dotted lines in Figure~\ref{weak_gamma_V}), in agreement with the fact that the only voltage dependency in Eqs.~(\ref{Seepp}) to (\ref{Shhpbarp}) comes from the factor in front of the $\mathcal{F}$ function. It is not more the case in the presence of interactions since they strongly affect the dot occupation number and thus the behavior of the mixed and heat correlators (see for example the black solid line in the upper right graph in Figure~\ref{weak_gamma_V}). In addition, we notice that: (i)~the mixed-correlators can change their sign, (ii)~$\mathcal{S}_{LR}^{eh}$ and $\mathcal{S}_{LR}^{he}$ are related to each other by changing $V$ in $-V$, and (iii)~the charge correlator $\mathcal{S}_{LL}^{ee}$ and the heat cross-correlator $\mathcal{S}_{LR}^{hh}$ are even functions of the bias voltage. These last two properties are however related to the particular case we have treated here, i.e. a symmetrically applied bias voltage $\mu_{L,R}=\pm eV/2$. Concerning the effect of Coulomb interactions, the main observation is that they have a strong influence on the correlators when the coupling $\Gamma$ is weak, but this effect is attenuated when $\Gamma$ increases as we can see on Figure~\ref{strong_gamma_V}. Indeed, when the coupling $\Gamma$ between the dot and the reservoirs increases, the spectral response of the central dot widens at energies $\epsilon_0$ and $\epsilon_0+U$, showing the loss of discreteness of the central system. As a consequence, at strong $\Gamma$, the effect of Coulomb interactions weakens, or inversely, at weak $\Gamma$, the effect of Coulomb interactions is stronger. Moreover, the additional structures brought by the Coulomb interactions disappear excepted in $\mathcal{S}_{LL}^{ee}$ wherein we can still see the intermediate plateau. Another difference between the weak and strong coupling regimes is the fact that in the strong coupling regime, the Coulomb interactions reduce or let almost unchanged the correlators whereas in the weak-coupling regime, the correlators can be either strongly enhanced or on the contrary reduced by the interactions according to the voltage value.

\begin{figure}
\includegraphics[width=12pc]{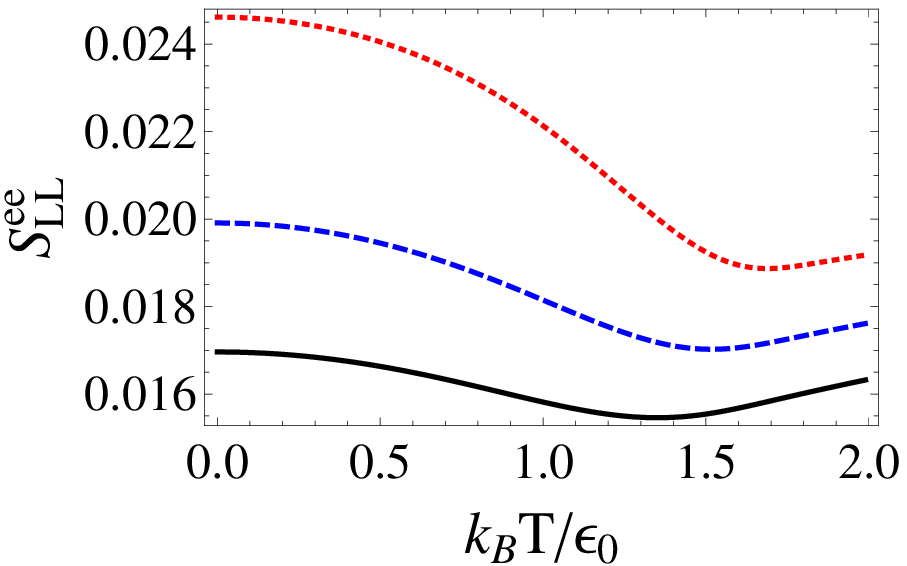}
\includegraphics[width=12pc]{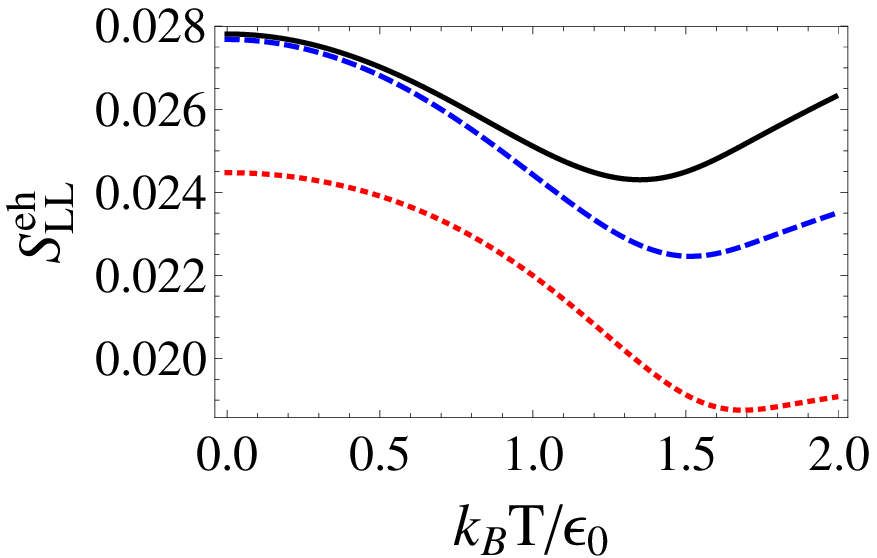}
\includegraphics[width=12pc]{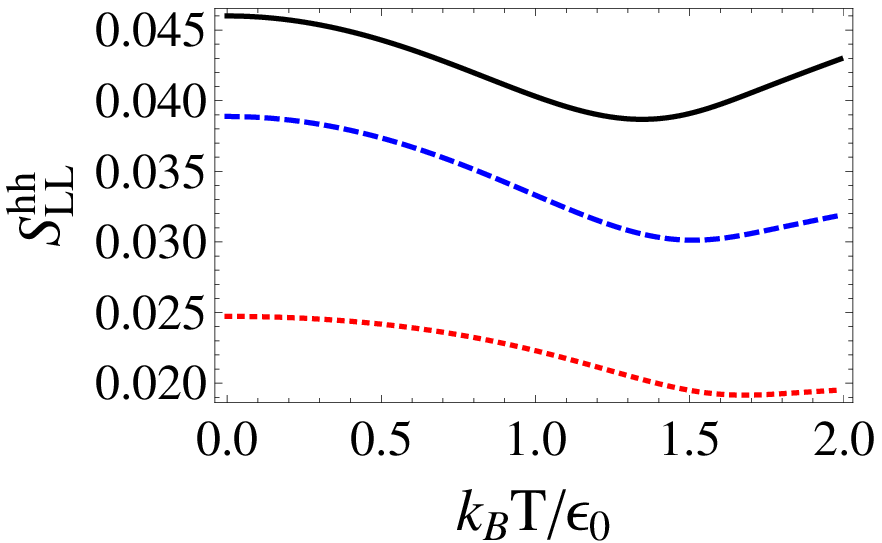}\\
\includegraphics[width=12pc]{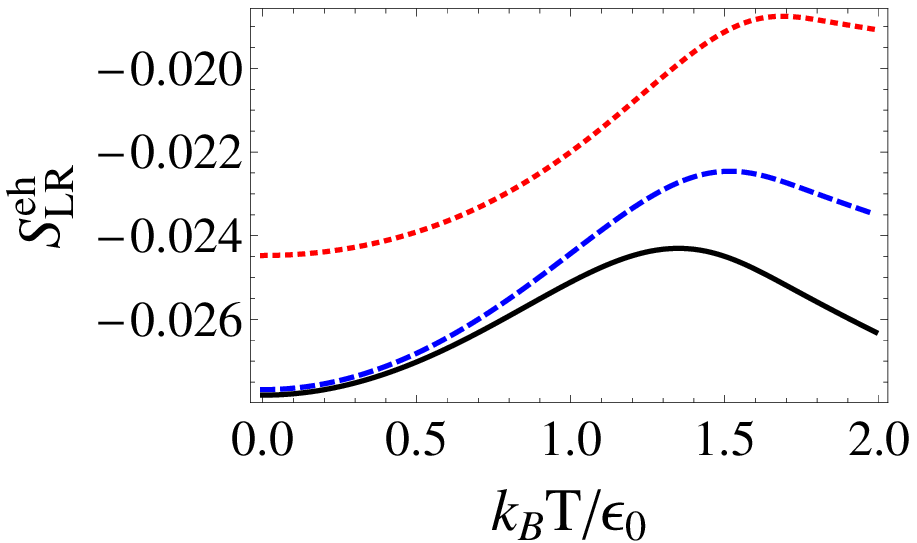}
\includegraphics[width=12pc]{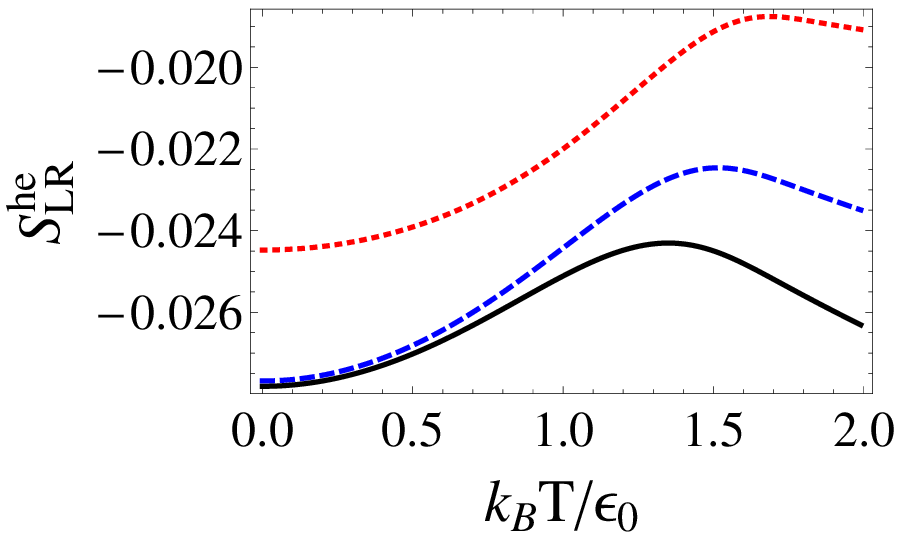}
\includegraphics[width=12pc]{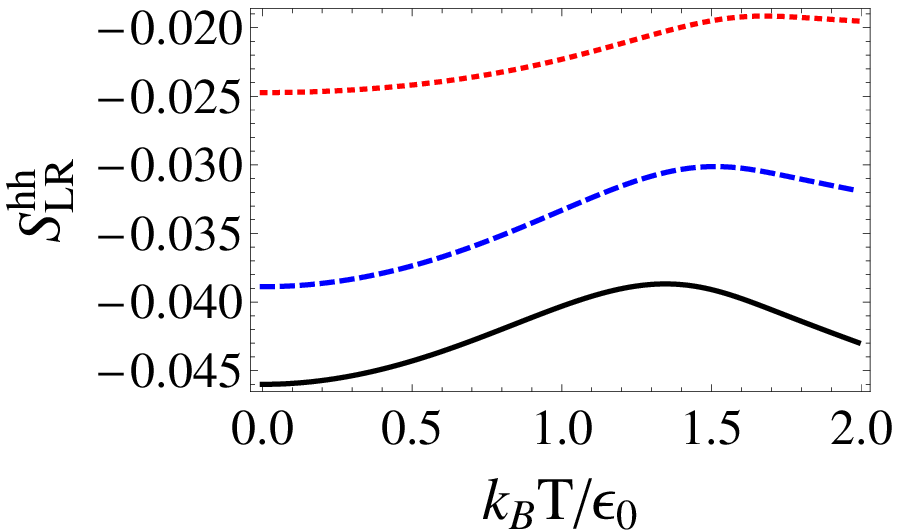}
\caption{\label{weak_gamma_T}Current correlators as a function of temperature at $\Gamma/\epsilon_0=0.01$, $k_BT_0/\epsilon_0=1$, and $V=0$, for $U=0$ (red dotted lines), $U/\epsilon_0=1$ (blue dashed lines) and $U/\epsilon_0=2$ (black solid lines).}
\end{figure}

\begin{figure}
\begin{center}
\includegraphics[width=12pc]{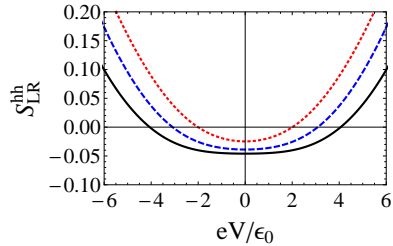}
\caption{\label{heat_cross_correlator}The heat cross-correlator as a function of voltage at $k_BT_0/\epsilon_0=1$ for $\Gamma/\epsilon_0=0.01$ and $T=0$, with $U=0$ (red dotted lines), $U/\epsilon_0=1$ (blue dashed lines) and $U/\epsilon_0=2$ (black solid lines).}
\end{center}
\end{figure}

\subsection{Current correlators as a function of the temperature gradient}

Figure~\ref{weak_gamma_T} shows the auto- and cross-correlators for a weak value of the coupling $\Gamma$ as a function of the temperature gradient for various values of the Coulomb interactions (compare the red dotted, blue dashed and black solid lines). The  bias voltage is set to zero. In that regime, $\mathcal{S}_{LR}^{eh}$ and $\mathcal{S}_{LR}^{he}$ coincide and $\mathcal{S}_{LR}^{hh}=-\mathcal{S}_{LL}^{hh}$ unlike to what happens in the low temperature limit of Figures~\ref{weak_gamma_V} and~\ref{strong_gamma_V}. Interestingly, whereas the Coulomb interactions decrease the value of the charge correlator, it increases the absolute value of the mixed and heat correlators when a temperature gradient is applied. 
Assuming that the mixed correlators give indication on the thermoelectric figure of merit, as demonstrated in the linear response regime~\cite{crepieux2014}, the Coulomb interactions are expected to affect the thermoelectric conversion. Beside, we notice that the variation of the correlators is not monotonous since a minimum is observed in their absolute values at a temperature gradient which decreases when $U$ increases. This is due to a partial compensation between the thermal and voltage contributions contained in Eq.~(\ref{F}). 
Another important result is the fact that the heat cross-correlator changes its sign between the configuration where average temperature $T_0$  is small and the configuration where average temperature $T_0$ is large (compare the right bottom graph of Figure~\ref{weak_gamma_V} in which the sign is positive to the right bottom graph of Figure~\ref{weak_gamma_T} in which the sign is negative). To improve our understanding of this effect, we plot $\mathrm{S}^{hh}_{LR}$ in Figure~\ref{heat_cross_correlator} at higher $T_0$ as a function of the voltage bias: the heat cross-correlator changes its sign and is not necessarily negative as the charge cross-correlator does. This result is in agreement with Ref.~\cite{moskalets2014}. The effect of the interactions is to broaden the domain of voltage where the heat cross-correlator is negative. 
%When $\Gamma$ is increased, the curves we obtain present similar characteristics, we decide not to show them.

\section{Conclusion}

In this work, we have shown that in the regime of weak coupling between the dot and the reservoirs, the Coulomb interactions can strongly affect the charge, heat and mixed current correlators: they introduces structures in their profiles when the bias is varied and they increase the amplitude of the correlators when the temperature is changed except for the charge auto-correlator. In fact, the charge auto-correlator differs from the other correlators: it presents distinct characteristic such as an identical value at high voltage whatever is the interaction strength, and as mentioned just above it also exhibits at high temperature a decrease of its amplitude when interactions increase, opposite to what happens for the other correlators. Another important feature that has been highlighted is the strongly non-linear and non-monotonous profiles of the correlators both in their variations with voltage and temperature, except once more for the charge auto-correlator which is linear in voltage in the strong coupling regime. Moreover, we have find that the heat cross-correlator can change its sign and be positive while the charge cross-correlator is always negative.

\section{Acknowledgments}

A.C. wishes to thank Mireille Lavagna for valuable discussions.

\end{document}